\documentclass[12pt]{article}
\pdfoutput=1
\usepackage{hyperref}
\usepackage{graphicx}
\usepackage{graphics}
\usepackage{dsfont}
\usepackage{epsfig}
\usepackage{amsmath,amssymb,amsthm,amscd}
\usepackage{simpler-wick}

\newcommand{\be}{\begin{equation}}
\newcommand{\ee}{\end{equation}}
\newcommand{\ba}{\begin{aligned}}
\newcommand{\ea}{\end{aligned}}

\numberwithin{equation}{section}

\begin{document}
\begin{titlepage}

\rightline{USTC-ICTS/PCFT-22-14}

\vskip 3 cm

\centerline{\Large 
\bf  
Modular   Anomaly Equation   } 
\vskip 0.2 cm
\centerline{\Large 
\bf  
 for Schur Index of $\mathcal{N}=4$ Super-Yang-Mills     }

\vskip 0.5 cm

\renewcommand{\thefootnote}{\fnsymbol{footnote}}
\vskip 30pt \centerline{ {\large \rm 
Min-xin Huang\footnote{minxin@ustc.edu.cn}  
} } \vskip .5cm  \vskip 20pt 

\begin{center}
{Interdisciplinary Center for Theoretical Study,  \\ \vskip 0.1cm  University of Science and Technology of China,  Hefei, Anhui 230026, China} 
 \\ \vskip 0.3 cm
{Peng Huanwu Center for Fundamental Theory,  \\ \vskip 0.1cm  Hefei, Anhui 230026, China} 
\end{center}

\setcounter{footnote}{0}
\renewcommand{\thefootnote}{\arabic{footnote}}
\vskip 40pt
\begin{abstract}

We propose a novel modular anomaly equation for the unflavored Schur index in the $\mathcal{N}=4$ $SU(N)$ super-Yang-Mills theory. The vanishing conditions overdetermine the modular ambiguity ansatz from the equation, thus together they are sufficient to recursively compute the exact Schur indices for all $SU(N)$ gauge groups. Using the representations as MacMahon's generalized sum-of-divisors functions and Jacobi forms, we then prove our proposal as well as elucidate a general formula conjectured by Pan and Peelaers. 

\end{abstract}

\end{titlepage}
\vfill \eject


\newpage

\baselineskip=16pt

\tableofcontents

\section{Introduction}

As a type of Witten index, the superconformal indices \cite{Kinney:2005ej} encode the BPS spectrum of the theory, and have been studied extensively in the literature. The case of $\mathcal{N}=4$  super-Yang-Mills theory with $SU(N)$ gauge group is particularly interesting due to the holographic duality with type IIB string theory on $AdS_5\times S^5$ background \cite{Maldacena:1997re}. The superconformal indices have many important applications. Most notably,  they are essential for the understandings of the microscopic entropy of supersymmetric $AdS_5$ black holes \cite{Cabo-Bizet:2018ehj, Choi:2018hmj, Benini:2018ywd}. Their various expansions can be interpreted as the contributions of D-branes, studied e.g. recently in \cite{Arai:2020qaj, Imamura:2021ytr, Gaiotto:2021xce, Murthy:2022ien, Honda:2022hvy}. 


For theories with a Lagrangian description, the $d$-dimensional superconformal index can be computed by path integral formalism as the supersymmetric partition function on $S^1 \times S^{d-1}$, which localizes to a matrix integral, see e.g. an early paper on the case of $\mathcal{N}=4$  super-Yang-Mills \cite{Sundborg:1999ue}. A particular specialization of the 4d superconformal index, known as the Schur index \cite{Gadde:2011uv}, has some further nice mathematical properties. For example, in some cases it can be computed from the q-deformed 2d Yang-Mills \cite{Gadde:2011ik}, or the vacuum character of a corresponding chiral algebra \cite{Beem:2013sza}. For the case of $\mathcal{N}=4$ supersymmetry, besides a universal fugacity parameter denoted as $q$, the Schur index may have an extra flavor fugacity from the symmetry $SU(2)_F\subset SU(4)_R$. In this paper we will simply consider the unflavored index without the extra fugacity. Some remarkable (quasi)-modular properties of the index are studied recently in \cite{Pan:2021mrw, Beem:2021zvt} in the context of a larger class of theories, based on some earlier works in e.g. \cite{Bourdier:2015wda, Kang:2021lic}.

On the other hand, topological string theory on Calabi-Yau three-folds has been an active research area for decades, with many sophisticated available techniques. The goal of the present paper is to apply one of these techniques to the calculations of Schur index. The relation between superconformal index and topological string amplitude has appeared before, in e.g. \cite{Kim:2012gu, Iqbal:2012xm}. In those cases, one has a 5d supersymmetric field theory from compactifying M-theory on a Calabi-Yau three-fold, and the 5d Nekrasov partition function on the Omega background $S^1\times \mathbb{R}^4_{\epsilon_1, \epsilon_2}$ is simply equivalent to the refined topological string amplitude on the Calabi-Yau space.  The 5d superconformal index at the fixed point of renormalization group flow can be computed by localization method as the partition function of the 5d field theory on $S^1\times S^4$, and is written as an integral of a product of two complex conjugate refined topological string amplitudes. This is similar to Pestun's calculation \cite{Pestun:2007rz} of $\mathcal{N}=2$ supersymmetric partition function on $S^4$, which localizes to a matrix integral in terms of 4d Nekrasov partition function. Similar relations appear also for 5d supersymmetric partition function on $S^5$ and 6d superconformal index, which are computed by an integral of a triple product of refined topological string amplitudes \cite{Lockhart:2012vp}. 

Our setting is somewhat different from those of \cite{Kim:2012gu, Iqbal:2012xm, Lockhart:2012vp}, as the 4d superconformal index considered here seems much simpler than the 5d or 6d cases. We will directly apply topological string method of modular anomaly equation to the calculations of 4d Schur index, instead of writing it as an integral of topological string amplitudes. We will encounter the  Eisenstein series and Jacobi Theta functions, where some of the basic properties are listed in Appendix \ref{appen}. It is well known that the Eisenstein series $E_4, E_6$ freely generate the modular forms of $SL(2,\mathbb{Z})$. The second Eisenstein series $E_2$ is not exactly modular but transforms with a shift. The ring of polynomials of $E_2, E_4, E_6$, known as quasi-modular forms, is closed under the derivative action $q\frac{d}{dq}$. For a general introduction see \cite{Zagierbook}. The quasi-modular forms appear in many studies in topological string theory, especially  in geometries containing  elliptic curves, e.g. in early papers \cite{10.1007/978-1-4612-4264-2_5, 10.1007/978-1-4612-4264-2_6, Minahan:1997ct, Minahan:1998vr}. In some cases there is a modular anomaly equation containing derivative with respect to the quasi-modular $E_2$, which is related to the holomorphic anomaly equation for general Calabi-Yau geometries without necessarily elliptic curves \cite{Bershadsky:1993cx}. See e.g. the recent papers \cite{Huang:2015sta, Huang:2020dbh} for more discussions. 

We will propose an  analogous  modular anomaly equation for Schur index in our context. During our study we will utilize the interesting connection to the seemingly remote topic of number theory through the MacMahon's sum-of-divisors functions, whose mathematical properties \cite{fcd25745b70a45868e37b97d86434d65, cite-key} provide a proof of our proposal as well as elucidate the connections with available results in the literature.

\section{Modular anomaly equation}

According to the localization method, the unflavored Schur index of the $\mathcal{N}=4$ $SU(N)$ super-Yang-Mills theory can be written in terms of a unitary matrix integral. As in the literature  \cite{Pan:2021mrw, Beem:2021zvt}, it is convenient to treat the even and odd ranks of the gauge groups separately. We consider first the simpler $SU(2N+1)$ case. The formula for Schur index is 
\be \label{indexodd}
\mathcal{I}_{2N+1} (q) =\frac{q^{\frac{N(N+1)}{2}}} {(2N+1)!}   \prod_{n=1}^{\infty}(\frac{1-q^{n-\frac{1}{2} } }{1-q^{n} })^2 \oint\prod_{i=1}^{2N+1 }  \frac{d z_i}{2\pi i z_i} \prod_{i\neq j} (1-\frac{z_i}{z_j}) \textrm{PE} [ i_V(q^{\frac{1}{2}} ) (\sum_{i,j=1}^{2N+1} \frac{z_i}{z_j} ) ], 
\ee
where $i_V(q) = \frac{2q}{1+q}$ is the $1/8$ BPS letter index, and PE denotes the well known plethystic exponential applied to all variables $q, z_i$. Here the factor  $\prod_{n=1}^{\infty}(\frac{1-q^{n-\frac{1}{2} } }{1-q^{n} })^2$ accounts for the difference between special unitary group and unitary group. We have also chosen the prefactor $q^{\frac{N(N+1)}{2}}$ in the convention so that the results would have nice modular properties.  For a finite $N$, it is not difficult to perform the contour integrals which are residues around $z_i\sim 0$ to obtain the $q$-expansion series to a finite order. For special unitary group, the integration variables would satisfy the product constraint $\prod_{i=1}^{2N+1} z_i =1$,  so we only need to do the first $2N$ contour integrals and the last variable $z_{2N+1}$ will automatically drop out. Although the formula appears to have half integer powers in the $q$-expansion, the result actually has only integer powers. From the formula (\ref{indexodd}) it is obvious that the $q$-expansion starts at a high power as 
\be \label{vanishingodd} 
\mathcal{I}_{2N+1} (q) =\mathcal{O} (q^{\frac{N(N+1)}{2}}) . 
\ee
 
The exact calculations of (\ref{indexodd}) were first performed in \cite{Bourdier:2015wda} in terms of elliptic integrals and there is also an all order $q$-series formula 
\be \label{qseries}
 \mathcal{I}_{2N+1} (q) =  \prod_{m=1}^{\infty}(1-q^{m} )^{-3} \sum_{n=0}^{\infty} (-1)^n [\binom{2N+1+n}{2N+1}+\binom{2N+n}{2N+1} ] q^{\frac{(n+N)(n+N+1)}{2}} .
\ee
The results were organized into nice formulas in terms of quasi-modular forms in \cite{Pan:2021mrw, Beem:2021zvt}.  We can list the formulas in term of Eisenstein series for the first few orders 
\be \ba  \label{Schurodd}
& \mathcal{I}_{1} (q) =1 , ~~~~~~ \mathcal{I}_{3} (q) = \frac{E_2} {2} +\frac{1}{24},  \\
& \mathcal{I}_{5} (q) =  \frac{E_2^2}{ 8} - \frac{E_4}{4} + \frac{E_2}{16} + \frac{3}{640},   \\
& \mathcal{I}_{7} (q) = \frac{E_2^3}{48}-\frac{E_2E_4}{8} +\frac{E_6}{6} +\frac{5 E_2^2}{192}  -\frac{5 E_4}{96}  +\frac{37 E_2}{3840} + \frac{5}{7168}. 
\ea \ee
A general formula for all $N$'s is also conjectured by Pan and Peelaers  \cite{Pan:2021mrw} as 
\be  \label{general} 
\mathcal{I}_{2N+1} = \sum_{k=0}^{N} \lambda^{(N)}_{k} \tilde{\mathbb{E}}_{2k}, 
\ee
where $\lambda^{(N)}_{k}$'s are constants determined by some rather complicated relations, and we will instead give a simpler recursion relation as well as an elementary generating function for computing them below.  $\tilde{\mathbb{E}}_{2k}$ is a quasi-modular form of homogeneous weight $2k$ defined by 
\be \label{weight}
 \tilde{\mathbb{E}}_{0}=1, ~~~~ \tilde{\mathbb{E}}_{2k} = \sum_{\sum_{j\geq 1} jn_j =k } ~\prod_{p\geq 1} \frac{1}{n_p!} (-\frac{E_{2p}}{2p})^{n_p}~ .
\ee
So the weight $2k$ component in the Schur index $\mathcal{I}_{2N+1}$ is universal, i.e. independent of $N$ up to a constant factor.

Inspired particularly by the studies of the BPS partition functions of E-strings in \cite{Minahan:1997ct}, we propose the following modular anomaly equation for the Schur index 
\be \label{MAE}
\partial_{E_2} \mathcal{I}_{2N+1} = \sum_{k=1}^N c_k \mathcal{I}_{2N+1-2k}, 
\ee
where $c_k$ are some constants to be determined in a moment. We note that by string duality, the partition function in \cite{Minahan:1997ct} is equivalent to genus zero sector of topological string theory on a local half K3 Calabi-Yau space, and the modular anomaly equation has been subsequently generalized to higher genus \cite{Hosono:1999qc} and to refined theory \cite{Huang:2013yta}. The modular anomaly equation in  \cite{Minahan:1997ct} is recursive in the number of E-strings, which is identified with the rank of gauge group in another equivalent description in terms of $\mathcal{N}=4$ topological Yang-Mills theories on a half K3 surface \cite{Minahan:1998vr}. Therefore it is reasonable that we can also have an equation (\ref{MAE}) recursive in the rank of the gauge group. There are certainly some notable differences with the usual form of modular anomaly equation familiar in topological string theory. First, the right hand side of our equation (\ref{MAE}) is purely linear in the lower rank indices, without the usual quadratic terms. Secondly, as seen from (\ref{Schurodd}), the Schur index is inhomogeneous, i.e. a combination of quasi-modular forms of different weights, unlike the usual homogenous forms. 

The modular anomaly equation (\ref{MAE}) determines the Schur index up to an $E_2$ independent term, a modular ambiguity which is polynomial of $E_4, E_6$. Since the index $\mathcal{I}_{2N+1}$ has a maximal weight of $2N$, the number of unknown coefficients in the ansatz for modular ambiguity can be easily counted. In general, the dimension of the space of modular forms of weight $2N$ is no more than $[\frac{N}{6}]+1$. So in our case we can estimate the number of unknown coefficients $\sum_{k=0}^N( [\frac{k}{6}]+1)\sim \frac{N^2}{12}$ for large $N$. On the other hand, for a generic modular ambiguity, the $q$-expansion of the Schur index starts from the lowest constant $q^0$ term. Similar to the case in \cite{Minahan:1998vr}, the vanishing condition (\ref{vanishingodd}) imposes very strong constrains, generically fixing $\frac{N(N+1)}{2}$ unknown coefficients, always overdetermining the ansatz. Staring from a very simple initial condition $ \mathcal{I}_{1} (q) =1, c_1=\frac{1}{2}$, we can recursively efficiently compute all Schur indices $\mathcal{I}_{2N+1}$ and also determine the constants $c_k$'s in (\ref{MAE}), which are $\frac{1}{2}, \frac{1}{24}, \frac{1}{180}, \frac{1}{1120}, \frac{1}{6300}, \cdots$. We then observe a general formula for the constants 
\be  \label{constants}
c_k = \frac{(k-1)!^2}{(2k)!} .
\ee

Our anomaly equation (\ref{MAE}) is compatible with the general formula (\ref{general}). It is easy to see that $\partial_{E_2}  \tilde{\mathbb{E}}_{2k+2} =-\frac{1}{2}  \tilde{\mathbb{E}}_{2k}$, so the weight $2k$ components of each term in (\ref{MAE}) are always proportional to $\tilde{\mathbb{E}}_{2k}$. More precisely, comparing the coefficients in (\ref{MAE}) and (\ref{general}) we find the relation 
\be  \label{eq10}
\lambda^{(N)}_{k+1} =-2 \sum_{l=1}^N c_l \lambda^{(N-l)}_k,~~~ k\geq 0. 
\ee

There is also another interesting method to compute the Schur index. It is pointed out in \cite{Kang:2021lic} that in this case, the Schur index is simply a MacMahon's generalized sum-of-divisors function
\be 
\mathcal{I}_{2N+1} (q) = \sum_{0<m_1<\cdots <m_N} \frac{q^{m_1+\cdots+m_N} }{(1-q^{m_1})^2 \cdots (1-q^{m_N})^2} .
\ee
In \cite{fcd25745b70a45868e37b97d86434d65}, a recursion relation for the  MacMahon's function is derived 
\be  \label{recur}
\mathcal{I}_{2N+1} (q) = \frac{1}{2N(2N+1)} [(6\mathcal{I}_{3} (q) +N(N-1) )\mathcal{I}_{2N-1} (q) -2q\frac{d}{dq} \mathcal{I}_{2N-1} (q) ]. 
\ee
Using the derivative relations of quasi-modular forms (\ref{Ramanujan}), it is the clear that $\mathcal{I}_{2N+1}$ is a inhomogeneous quasi-modular form of weight $2N$, and it can be also easily computed recursively. The $q$-series formula (\ref{qseries}) was also proved in \cite{fcd25745b70a45868e37b97d86434d65}, therefore the equivalence of Schur index and MacMahon's function in this case is clear. The structure of formula (\ref{general}) of Schur index is preserved by the recursion (\ref{recur}) due to the following derivative formula 
\be 
q\frac{d}{dq}  \tilde{\mathbb{E}}_{2k-2} = k(2k+1)  \tilde{\mathbb{E}}_{2k} -3  \tilde{\mathbb{E}}_2 \tilde{\mathbb{E}}_{2k-2} , 
\ee
which is a generalization of Ramanujan formulas (\ref{Ramanujan}) and can be certainly checked for any finite $k$. It should be derivable from the differential equations of the twisted Eisenstein series used in  \cite{Pan:2021mrw}. The  relation (\ref{recur}) is then equivalent to a recursion for the coefficients 
\be \label{recursion15}
\lambda^{(N)}_k =\frac{1}{8N(2N+1)}[ (2N-1)^2 \lambda^{(N-1)}_k -8k(2k+1)\lambda^{(N-1)}_{k-1}],
\ee
where in the derivation we only need to look at the $E_2$ monomial term in $\tilde{\mathbb{E}}_{2k}$ in (\ref{weight}). For $k<0$ or $k>N$ the coefficients are defined as  $\lambda^{(N)}_k=0$. From a simple initial condition $\lambda^{(0)}_0 =1$ we can then use the recursion (\ref{recursion15}) to compute all coefficients. For the special cases $k=0$ or $k=N$, simple formulas $\lambda^{(N)}_0 =\frac{(2N)!}{2^{4N} (2N+1)N!^2}$ and $\lambda^{(N)}_N=(-1)^N$ can be easily derived from the recursion. The recursion (\ref{recursion15}) looks much simpler than those given in \cite{Pan:2021mrw} but they should certainly give the same result.

In the paper \cite{cite-key}, Rose further considered more general MacMahon's sum-of-divisors functions, and provide formulas for the generating functions in terms of Jacobi forms. A key ingredient in the proofs of the formulas in \cite{fcd25745b70a45868e37b97d86434d65, cite-key} is the well known Jacobi triple product identity. For an introduction of Jacobi forms, see \cite{eichler2013theory, Dabholkar:2012nd}. This turns out to provide a proof of the anomaly equation (\ref{MAE}). In our case, the generating function for Schur index can be written in terms of the Jacobi theta function as
\be  \label{eq2.14}
F(q, x):= \sum_{N=0}^{\infty} (-1)^N \mathcal{I}_{2N+1}(q)  x^{2N+1} = \frac{i\theta_1(q, z)} {\eta(q)^3} , 
\ee
with identification of parameters $x=e^{\pi i z} -e^{-\pi i z}$.  It is known that a Jacobi forms $\phi_m$ of index $m$ satisfies a modular anomaly equation
\be 
(\partial_{E_2} -m (2\pi z)^2 ) \phi_m =0. 
\ee
This has been applied successfully in topological string theory for making ansatz, see e.g. \cite{Huang:2015sta}. In our context,  the generating function is not exactly a Jacobo form of $SL(2, \mathbb{Z})$, but of a subgroup with index $\frac{1}{2}$ \cite{cite-key}. The modular anomaly equation can be still applied similarly
\be  
(\partial_{E_2} -\frac{1}{2} (2\pi z)^2 ) F(q, x) =0. 
\ee
Using the relation  $x=e^{\pi i z} -e^{-\pi i z}$ as mentioned below (\ref{eq2.14}), we can solve for the inverse relation 
\be  
\pi i z =  \textrm{arcsinh} (\frac{x}{2}) =  \log [\frac{1}{2} (x+\sqrt{4+x^2})] . 
\ee
Denoting $f(x):=  (\pi z)^2 $, it is easy to check that $f(x)$ satisfies a differential equation $(x^2+4) f^{\prime\prime} (x)+xf^{\prime} (x) + 2 =0$. So we can straightforwardly prove by induction that it  has the following series expansion 
\be  \label{fx}
f(x) = - \log^2[\frac{1}{2} (x+\sqrt{4+x^2})] = \frac{1}{2} \sum_{n=1}^{\infty} (-1)^n \frac{(n-1)!^2}{(2n)!} x^{2n}. 
\ee
Thus we have derived the modular anomaly equation (\ref{MAE}) with the formulas (\ref{constants}) for the coefficients.

We can define a generating function $G(x,y) := \sum_{N=0}^{\infty}\sum_{k=0}^N \lambda^{(N)}_k x^{2N+1} y^{2k+1}$.  Using the relation (\ref{eq10}), we have 
\be 
G(x,y) +2y^2 G(x,y) f(ix) = \sum_{N=0}^{\infty} \frac{(2N)!}{2^{4N} (2N+1)N!^2} x^{2N+1} y = 2y f(ix)^{\frac{1}{2}} . 
\ee
So we can also get a solution in terms of elementary functions 
\be \label{gen1}
G(x,y) =  \frac{ 2y f(ix)^{\frac{1}{2}} } {1+4y^2 f(ix)} .
\ee
One can check the recursion (\ref{recursion15}) is satisfied due to the differential equation 
\be  \label{differ}
[4\partial_x^2  - (x\partial_x)^2  + 4y^2 \partial_y^2 y^2 ] G(x,y)=0.
\ee

\section{The $SU(2N)$ case} 
Next we consider the $SU(2N)$ case, which is a little more complicated but similar. The Schur index formula in our convention is 
\be \label{indexeven}
\mathcal{I}_{2N} (q) =\frac{q^{\frac{N^2}{2}}} {(2N)!}   \oint\prod_{i=1}^{2N}  \frac{d z_i}{2\pi i z_i} \prod_{i\neq j} (1-\frac{z_i}{z_j}) \textrm{PE} [ i_V(q^{\frac{1}{2}} ) (\sum_{i,j=1}^{2N} \frac{z_i}{z_j} ) ], 
\ee
similar to (\ref{indexodd}) but with a different prefactor. We have omitted the factor  $\prod_{n=1}^{\infty}(\frac{1-q^{n-\frac{1}{2} } }{1-q^{n} })^2$ so that the expression would have better modular property in this case. So strictly speaking this is a ``rescaled" Schur index, but for convenience of notation we simply work with this better definition  in our context. The vanishing constrains for the index is 
\be \label{vanishingeven} 
\mathcal{I}_{2N} (q) =\mathcal{O} (q^{\frac{N^2}{2}}) . 
\ee
In this case, the $q$-expansion has half integer powers, so this generically will impose $N^2$ constrains on the ansatz. In this case the $q$-series expansion in \cite{Bourdier:2015wda} is 
\be  \label{qSU(2N)} 
 \mathcal{I}_{2N} (q) =  \prod_{m=1}^{\infty}\frac{1+q^{\frac{m}{2}}}{1- q^{\frac{m}{2}} } 
 \sum_{n=0}^{\infty} (-1)^n [\binom{2N+n}{2N}+\binom{2N+n-1}{2N} ] q^{\frac{(n+N)^2}{2}} .
\ee

The modular group is now $\Gamma^0(2)$, whose modular forms are generated by 
\be 
\Theta_{r,s} (q) = \theta_2(q) ^{4r} \theta_3(q) ^{4s} + \theta_2(q) ^{4s} \theta_3(q) ^{4r}, 
 \ee
which has weight $2(r+s)$. Some low order formulas for the Schur indices are also available in  \cite{Pan:2021mrw, Beem:2021zvt}
\be \ba  \label{Schureven}
&  \mathcal{I}_{2} (q) = \frac{E_2} {2} +\frac{\Theta_{0,1}}{24},  \\
& \mathcal{I}_{4} (q) = \frac{E_2^2}{8} + \frac{E_2 \Theta_{0,1} }{48}     + \frac{\Theta_{0, 2}}{1152}   - \frac{\Theta_{1, 1}}{576}    +\frac{E_2}{24} +  \frac{\Theta_{0, 1}}{288} .
\ea \ee
Similarly, in this case we propose the modular anomaly equation 
\be \label{MAEeven}
\partial_{E_2} \mathcal{I}_{2N} = \sum_{k=1}^N c_k \mathcal{I}_{2N-2k}, 
\ee
with a convention for initial index $\mathcal{I}_0=1$. The number of unknown coefficients in the modular ambiguity in $\mathcal{I}_{2N}$ is counted by $\Theta_{r,s} (q)$'s with $r+s\leq N, r\leq s$, and goes like $\frac{N^2}{4}$ for large $N$, much smaller than the number of constrains $N^2$. It also turns out that there is no weight zero constant term in the modular ambiguity, as can be seen from the examples in (\ref{Schureven}). So starting also from the simple initial condition 
$\mathcal{I}_0=1, c_1=\frac{1}{2}$, we can compute all Schur indices and fix the constants $c_k$'s which turn out to be the same as in the $SU(2N+1)$ case (\ref{constants}). Of course we can also include the constant term in the ansatz for modular ambiguity, then we simply require the extra initial conditions for $\mathcal{I}_2, c_2$ to start the recursive algorithm.

The Schur index can be represented by another MacMahon's generalized sum-of-divisors function appeared in \cite{fcd25745b70a45868e37b97d86434d65} as
\be 
\mathcal{I}_{2N} (q) = \sum_{0<m_1<\cdots <m_N} \frac{q^{m_1+\cdots+m_N-\frac{N}{2}} }{(1-q^{m_1-\frac{1}{2}})^2 \cdots (1-q^{m_N-\frac{1}{2} })^2} ,
\ee
where the same series expansion (\ref{qSU(2N)}) was also derived. Furthermore a recursion relation is proved in \cite{fcd25745b70a45868e37b97d86434d65} 
\be  \label{recur24}
\mathcal{I}_{2N} = \frac{1}{2N(2N-1)} [(2\mathcal{I}_{2} +(N-1)^2) \mathcal{I}_{2N-2} -2q\frac{d}{dq} \mathcal{I}_{2N-2}] .
\ee
From the recursion and the initial formula $\mathcal{I}_2$, it is clear that the Schur index $\mathcal{I}_{2N}$ is an inhomogeneous  $\Gamma^0(2)$ quasi-modular form of weight $2N$. A general formula is also conjectured in \cite{Pan:2021mrw} in this case. In our convention it is 
\be \ba \label{general25}
 \mathcal{I}_{2N} =  \sum_{k=0}^N  \frac{\tilde{\lambda}^{(N)}_k }{k!(2k-1)!!} \theta_4^{-1} (q\frac{d}{d q})^k \theta_4    
  = \sum_{k=0}^N  \frac{\tilde{\lambda}^{(N)}_k }{k!} (-\frac{E_2}{2})^k +\cdots,
\ea \ee 
where analogous to the $SU(2N+1)$ case in the previous section, we denote the coefficients of $E_2$ monomial terms with the same prefactor.  It is easy to check that the recursion  (\ref{recur24}) preserves the structure the general formula (\ref{general25}) using derivative formula $ q\frac{d}{d q} \log \theta_4 = -\mathcal{I}_2$. Furthermore it provides a recursion for the coefficients 
\be  \label{recureven}
\tilde{\lambda}^{(N)}_k =\frac{1}{2N(2N-1)} [(N-1)^2 \tilde{\lambda}^{(N-1)}_k -2k(2k-1)\tilde{\lambda}^{(N-1)}_{k-1} ] .
\ee
From the initial condition $\tilde{\lambda}^{(0)}_0=1$ we can compute all coefficients (again $\tilde{\lambda}^{(N)}_k=0$ for $k<0$ or $k>N$). In this case it is easy to see $\tilde{\lambda}^{(N)}_0 =0$ for all $N\geq 1$, which explain the absence of constant term in the Schur index observed earlier. The other special formula $\tilde{\lambda}^{(N)}_N =(-1)^N$ is the same as in the previous section. Finally from the anomaly equation (\ref{MAEeven}) we also have the relation 
\be \label{eq27} 
\tilde{\lambda}^{(N)}_{k+1} =-2 \sum_{l=1}^N c_l \tilde{\lambda}^{(N-l)}_k. 
\ee
Similarly the Jacobi form formula in this case \cite{cite-key}  provides a proof of the anomaly equation (\ref{MAEeven}), and we skip the details here.

Similarly we can define the generating function  $\tilde{G}(x,y) := \sum_{N=0}^{\infty}\sum_{k=0}^N \tilde{\lambda}^{(N)}_k x^{2N} y^{2k}$. The anomaly equation (\ref{eq27}) then provides a solution 
\be \label{gen2}
\tilde{G}(x,y)  =\frac{1}{1+4y^2 f(ix)}, 
\ee
where $f(x)$ is the same function (\ref{fx}) appeared before. The recursion (\ref{recureven}) is satisfied since  $\tilde{G}(x,y)$ is annihilated by the same differential operator as in (\ref{differ}).

\section{Discussions}

Although the results for Schur index in the current study have been available in the literature, we find our method of using the anomaly equation (\ref{MAE}) and the vanishing conditions (\ref{vanishingodd}) provides so far the simplest approach with minimal assumptions. The vanishing conditions are in fact highly redundant, providing consistency checks by themselves and automatically giving the coefficients (\ref{constants}) in the anomaly equation.  Furthermore, using the anomaly equation we are able to solve the generating functions (\ref{gen1}, \ref{gen2}) for the coefficients in the general formulas (\ref{general}, \ref{general25}) conjectured in \cite{Pan:2021mrw}. 

The non-trivial existence of such over-constrained systems likely suggests a natural underlying geometric explanation, as mentioned in \cite{fcd25745b70a45868e37b97d86434d65, cite-key}. It is would also be interesting to check whether the more general MacMahon's sum-of-divisors functions studied in \cite{cite-key} have connections with Schur indices of some other superconformal field theories. 

Our anomaly equation (\ref{MAE}) seems universally simple that it should have a wider applicability. It would be interesting to apply our proposal to more general superconformal indices, including more flavor fugacities. A better understanding of the modular property would help the analysis of the asymptotic behavior of the index, which is essential for accounting for the black hole entropy in holographic duality.

\vspace{0.2in} {\leftline {\bf Acknowledgments}}
\nopagebreak

We thank Bao-ning Du, Masazumi Honda, Jun-Hao Li, Yiwen Pan, Gao-fu Ren, Pei-xuan Zeng for helpful discussions and correspondences, Sheldon Katz, Albrecht Klemm, Kimyeong Lee, Yuji Sugimoto, Xin Wang for collaborations on related topics. This work was supported in parts by the national Natural Science Foundation of China (Grants  No.11947301 and No.12047502).

\appendix

\section{Eisenstein series and Jacobi theta functions}
\label{appen}

We use the following convention for the weight $2k$ Eisenstein series
\be
E_{2k} = - \frac{B_{2k}}{ (2k)!} + \frac{2}{(2k-1)!} \sum_{n=1}^{\infty} \frac{n^{2k-1} q^{n} }{ 1-q^{n}} .
\ee
The well known derivative formulas are due to Ramanujan 
\be \ba \label{Ramanujan}
& q\frac{d}{dq} E_2 = -E_2^2 +5E_4, ~~~~  q\frac{d}{dq} E_4 = -4E_2E_4 +14E_6,  \\
& q\frac{d}{dq} E_6 = -6E_2E_6 +\frac{60 E_4^2}{7}. 
\ea \ee
In the modular anomaly equation we need to take derivative with respect to $E_2$. Sometimes a commutation relation between the derivative actions is potentially helpful, see e.g. \cite{Huang:2020dbh}. For a homogeneous quasi-modular form $G_k$ of weight $k$, in the current convention for $E_2$ we have 
\be 
\partial_{E_2} q\frac{d}{dq} G_k = ( q\frac{d}{dq}  \partial_{E_2}-k) G_k.
\ee

The Jacobi theta function is defined by 
\be 
\theta\left[a \atop b\right](\tau,z)=\sum_{n\in \mathbb{Z}} e^{\pi i (n + a)^2 \tau + 2 \pi i z (n+a) + 2 \pi i n b}\  ,
\ee
with $q=e^{2\pi i \tau}$ and the usual auxiliary theta functions are $\theta_1= -i\theta\left[\frac{1}{2} \atop \frac{1}{2}\right]$, 
$\theta_2=\theta\left[\frac{1}{2} \atop 0\right]$, $\theta_3=\theta\left[0 \atop 0\right]$ and 
$\theta_4=\theta\left[0 \atop \frac{1}{2}  \right]$. 
Often we set the elliptic parameter $z=0$ and denote  
\be 
\theta_2 (q) = \sum_{n=-\infty}^{\infty} q^{\frac{1}{2} (n+\frac{1}{2})^2}, ~~~
\theta_3 (q) = \sum_{n=-\infty}^{\infty} q^{\frac{n^2}{2} }, ~~~   \theta_4 (q) = \sum_{n=-\infty}^{\infty} (-1)^n q^{\frac{n^2}{2} }. 
\ee
Then there is a relation $\theta_3^4 =\theta_2^4+\theta_4^4$, and the derivative formulas
\be \ba
& q\frac{d}{dq} \log\theta_2 = -\frac{1}{2} E_2 +\frac{1}{24}(\theta_3^4+\theta_4^4), ~~~
q\frac{d}{dq} \log\theta_3 = -\frac{1}{2} E_2 +\frac{1}{24}(\theta_2^4-\theta_4^4),  \\
& q\frac{d}{dq} \log\theta_4 = -\frac{1}{2} E_2 -\frac{1}{24}(\theta_2^4+\theta_3^4). 
\ea \ee

\addcontentsline{toc}{section}{References}


\begin{thebibliography}{10}

\bibitem{Kinney:2005ej}
J.~Kinney, J.~M. Maldacena, S.~Minwalla, and S.~Raju, ``{An Index for 4
  dimensional super conformal theories},''
  \href{http://dx.doi.org/10.1007/s00220-007-0258-7}{{\em Commun. Math. Phys.}
  {\bfseries 275} (2007) 209--254},
  \href{http://arxiv.org/abs/hep-th/0510251}{{\ttfamily arXiv:hep-th/0510251}}.

\bibitem{Maldacena:1997re}
J.~M. Maldacena, ``{The Large N limit of superconformal field theories and
  supergravity},'' \href{http://dx.doi.org/10.1023/A:1026654312961}{{\em Adv.
  Theor. Math. Phys.} {\bfseries 2} (1998) 231--252},
  \href{http://arxiv.org/abs/hep-th/9711200}{{\ttfamily arXiv:hep-th/9711200}}.

\bibitem{Cabo-Bizet:2018ehj}
A.~Cabo-Bizet, D.~Cassani, D.~Martelli, and S.~Murthy, ``{Microscopic origin of
  the Bekenstein-Hawking entropy of supersymmetric AdS$_{5}$ black holes},''
  \href{http://dx.doi.org/10.1007/JHEP10(2019)062}{{\em JHEP} {\bfseries 10}
  (2019) 062}, \href{http://arxiv.org/abs/1810.11442}{{\ttfamily
  arXiv:1810.11442 [hep-th]}}.

\bibitem{Choi:2018hmj}
S.~Choi, J.~Kim, S.~Kim, and J.~Nahmgoong, ``{Large AdS black holes from
  QFT},'' \href{http://arxiv.org/abs/1810.12067}{{\ttfamily arXiv:1810.12067
  [hep-th]}}.

\bibitem{Benini:2018ywd}
F.~Benini and P.~Milan, ``{Black Holes in 4D $\mathcal{N}$=4 Super-Yang-Mills
  Field Theory},'' \href{http://dx.doi.org/10.1103/PhysRevX.10.021037}{{\em
  Phys. Rev. X} {\bfseries 10} no.~2, (2020) 021037},
  \href{http://arxiv.org/abs/1812.09613}{{\ttfamily arXiv:1812.09613
  [hep-th]}}.

\bibitem{Arai:2020qaj}
R.~Arai, S.~Fujiwara, Y.~Imamura, and T.~Mori, ``{Schur index of the ${\cal
  N}=4$ $U(N)$ supersymmetric Yang-Mills theory via the AdS/CFT
  correspondence},'' \href{http://dx.doi.org/10.1103/PhysRevD.101.086017}{{\em
  Phys. Rev. D} {\bfseries 101} no.~8, (2020) 086017},
  \href{http://arxiv.org/abs/2001.11667}{{\ttfamily arXiv:2001.11667
  [hep-th]}}.

\bibitem{Imamura:2021ytr}
Y.~Imamura, ``{Finite-N superconformal index via the AdS/CFT correspondence},''
  \href{http://dx.doi.org/10.1093/ptep/ptab141}{{\em PTEP} {\bfseries 2021}
  no.~12, (2021) 123B05}, \href{http://arxiv.org/abs/2108.12090}{{\ttfamily
  arXiv:2108.12090 [hep-th]}}.

\bibitem{Gaiotto:2021xce}
D.~Gaiotto and J.~H. Lee, ``{The Giant Graviton Expansion},''
  \href{http://arxiv.org/abs/2109.02545}{{\ttfamily arXiv:2109.02545
  [hep-th]}}.

\bibitem{Murthy:2022ien}
S.~Murthy, ``{Unitary matrix models, free fermion ensembles, and the giant
  graviton expansion},'' \href{http://arxiv.org/abs/2202.06897}{{\ttfamily
  arXiv:2202.06897 [hep-th]}}.

\bibitem{Honda:2022hvy}
M.~Honda and T.~Yoda, ``{String theory, $\mathcal{N}=4$ SYM and Riemann
  hypothesis},'' \href{http://arxiv.org/abs/2203.17091}{{\ttfamily
  arXiv:2203.17091 [hep-th]}}.

\bibitem{Sundborg:1999ue}
B.~Sundborg, ``{The Hagedorn transition, deconfinement and N=4 SYM theory},''
  \href{http://dx.doi.org/10.1016/S0550-3213(00)00044-4}{{\em Nucl. Phys. B}
  {\bfseries 573} (2000) 349--363},
  \href{http://arxiv.org/abs/hep-th/9908001}{{\ttfamily arXiv:hep-th/9908001}}.

\bibitem{Gadde:2011uv}
A.~Gadde, L.~Rastelli, S.~S. Razamat, and W.~Yan, ``{Gauge Theories and
  Macdonald Polynomials},''
  \href{http://dx.doi.org/10.1007/s00220-012-1607-8}{{\em Commun. Math. Phys.}
  {\bfseries 319} (2013) 147--193},
  \href{http://arxiv.org/abs/1110.3740}{{\ttfamily arXiv:1110.3740 [hep-th]}}.

\bibitem{Gadde:2011ik}
A.~Gadde, L.~Rastelli, S.~S. Razamat, and W.~Yan, ``{The 4d Superconformal
  Index from q-deformed 2d Yang-Mills},''
  \href{http://dx.doi.org/10.1103/PhysRevLett.106.241602}{{\em Phys. Rev.
  Lett.} {\bfseries 106} (2011) 241602},
  \href{http://arxiv.org/abs/1104.3850}{{\ttfamily arXiv:1104.3850 [hep-th]}}.

\bibitem{Beem:2013sza}
C.~Beem, M.~Lemos, P.~Liendo, W.~Peelaers, L.~Rastelli, and B.~C. van Rees,
  ``{Infinite Chiral Symmetry in Four Dimensions},''
  \href{http://dx.doi.org/10.1007/s00220-014-2272-x}{{\em Commun. Math. Phys.}
  {\bfseries 336} no.~3, (2015) 1359--1433},
  \href{http://arxiv.org/abs/1312.5344}{{\ttfamily arXiv:1312.5344 [hep-th]}}.

\bibitem{Pan:2021mrw}
Y.~Pan and W.~Peelaers, ``{The exact Schur index in closed form},''
  \href{http://arxiv.org/abs/2112.09705}{{\ttfamily arXiv:2112.09705
  [hep-th]}}.

\bibitem{Beem:2021zvt}
C.~Beem, P.~Singh, and S.~S. Razamat, ``{Schur indices of class S and
  quasimodular forms},''
  \href{http://dx.doi.org/10.1103/PhysRevD.105.085009}{{\em Phys. Rev. D}
  {\bfseries 105} no.~8, (2022) 085009},
  \href{http://arxiv.org/abs/2112.10715}{{\ttfamily arXiv:2112.10715
  [hep-th]}}.

\bibitem{Bourdier:2015wda}
J.~Bourdier, N.~Drukker, and J.~Felix, ``{The exact Schur index of
  $\mathcal{N}=4$ SYM},'' \href{http://dx.doi.org/10.1007/JHEP11(2015)210}{{\em
  JHEP} {\bfseries 11} (2015) 210},
  \href{http://arxiv.org/abs/1507.08659}{{\ttfamily arXiv:1507.08659
  [hep-th]}}.

\bibitem{Kang:2021lic}
M.~J. Kang, C.~Lawrie, and J.~Song, ``{Infinitely many 4D N=2 SCFTs with a=c
  and beyond},'' \href{http://dx.doi.org/10.1103/PhysRevD.104.105005}{{\em
  Phys. Rev. D} {\bfseries 104} no.~10, (2021) 105005},
  \href{http://arxiv.org/abs/2106.12579}{{\ttfamily arXiv:2106.12579
  [hep-th]}}.

\bibitem{Kim:2012gu}
H.-C. Kim, S.-S. Kim, and K.~Lee, ``{5-dim Superconformal Index with Enhanced
  En Global Symmetry},'' \href{http://dx.doi.org/10.1007/JHEP10(2012)142}{{\em
  JHEP} {\bfseries 10} (2012) 142},
  \href{http://arxiv.org/abs/1206.6781}{{\ttfamily arXiv:1206.6781 [hep-th]}}.

\bibitem{Iqbal:2012xm}
A.~Iqbal and C.~Vafa, ``{BPS Degeneracies and Superconformal Index in Diverse
  Dimensions},'' \href{http://dx.doi.org/10.1103/PhysRevD.90.105031}{{\em Phys.
  Rev. D} {\bfseries 90} no.~10, (2014) 105031},
  \href{http://arxiv.org/abs/1210.3605}{{\ttfamily arXiv:1210.3605 [hep-th]}}.

\bibitem{Pestun:2007rz}
V.~Pestun, ``{Localization of gauge theory on a four-sphere and supersymmetric
  Wilson loops},'' \href{http://dx.doi.org/10.1007/s00220-012-1485-0}{{\em
  Commun. Math. Phys.} {\bfseries 313} (2012) 71--129},
  \href{http://arxiv.org/abs/0712.2824}{{\ttfamily arXiv:0712.2824 [hep-th]}}.

\bibitem{Lockhart:2012vp}
G.~Lockhart and C.~Vafa, ``{Superconformal Partition Functions and
  Non-perturbative Topological Strings},''
  \href{http://dx.doi.org/10.1007/JHEP10(2018)051}{{\em JHEP} {\bfseries 10}
  (2018) 051}, \href{http://arxiv.org/abs/1210.5909}{{\ttfamily arXiv:1210.5909
  [hep-th]}}.

\bibitem{Zagierbook}
D.~Zagier, \href{http://dx.doi.org/10.1007/978-3-540-74119-0_1}{``Elliptic
  modular forms and their applications,''} in {\em The 1-2-3 of modular forms},
  Universitext, pp.~1--103.
\newblock Springer, Berlin, 2008.
\newblock \url{https://doi.org/10.1007/978-3-540-74119-0_1}.

\bibitem{10.1007/978-1-4612-4264-2_5}
R.~Dijkgraaf, ``Mirror symmetry and elliptic curves,'' in {\em The Moduli Space
  of Curves}, R.~H. Dijkgraaf, C.~F. Faber, and G.~B.~M. van~der Geer, eds.,
  pp.~149--163.
\newblock Birkh{\"a}user Boston, Boston, MA, 1995.

\bibitem{10.1007/978-1-4612-4264-2_6}
M.~Kaneko and D.~Zagier, ``A generalized jacobi theta function and quasimodular
  forms,'' in {\em The Moduli Space of Curves}, R.~H. Dijkgraaf, C.~F. Faber,
  and G.~B.~M. van~der Geer, eds., pp.~165--172.
\newblock Birkh{\"a}user Boston, Boston, MA, 1995.

\bibitem{Minahan:1997ct}
J.~A. Minahan, D.~Nemeschansky, and N.~P. Warner, ``{Partition functions for
  BPS states of the noncritical E(8) string},''
  \href{http://dx.doi.org/10.4310/ATMP.1997.v1.n1.a7}{{\em Adv. Theor. Math.
  Phys.} {\bfseries 1} (1998) 167--183},
  \href{http://arxiv.org/abs/hep-th/9707149}{{\ttfamily arXiv:hep-th/9707149}}.

\bibitem{Minahan:1998vr}
J.~A. Minahan, D.~Nemeschansky, C.~Vafa, and N.~P. Warner, ``{E strings and N=4
  topological Yang-Mills theories},''
  \href{http://dx.doi.org/10.1016/S0550-3213(98)00426-X}{{\em Nucl. Phys. B}
  {\bfseries 527} (1998) 581--623},
  \href{http://arxiv.org/abs/hep-th/9802168}{{\ttfamily arXiv:hep-th/9802168}}.

\bibitem{Bershadsky:1993cx}
M.~Bershadsky, S.~Cecotti, H.~Ooguri, and C.~Vafa, ``{Kodaira-Spencer theory of
  gravity and exact results for quantum string amplitudes},''
  \href{http://dx.doi.org/10.1007/BF02099774}{{\em Commun. Math. Phys.}
  {\bfseries 165} (1994) 311--428},
  \href{http://arxiv.org/abs/hep-th/9309140}{{\ttfamily arXiv:hep-th/9309140}}.

\bibitem{Huang:2015sta}
M.-x. Huang, S.~Katz, and A.~Klemm, ``{Topological String on elliptic CY
  3-folds and the ring of Jacobi forms},''
  \href{http://dx.doi.org/10.1007/JHEP10(2015)125}{{\em JHEP} {\bfseries 10}
  (2015) 125}, \href{http://arxiv.org/abs/1501.04891}{{\ttfamily
  arXiv:1501.04891 [hep-th]}}.

\bibitem{Huang:2020dbh}
M.-x. Huang, S.~Katz, and A.~Klemm, ``{Towards refining the topological strings
  on compact Calabi-Yau 3-folds},''
  \href{http://dx.doi.org/10.1007/JHEP03(2021)266}{{\em JHEP} {\bfseries 03}
  (2021) 266}, \href{http://arxiv.org/abs/2010.02910}{{\ttfamily
  arXiv:2010.02910 [hep-th]}}.
  
  
  \bibitem{fcd25745b70a45868e37b97d86434d65}
G.~Andrews and S.~Rose, ``Macmahon's sum-of-divisors functions, chebyshev
  polynomials, and quasi-modular forms,''
  \href{http://dx.doi.org/10.1515/CRELLE.2011.179}{{\em Journal fur die Reine
  und Angewandte Mathematik} {\bfseries 2013} no.~676, (Jan., 2013) 97--103},
  \href{http://arxiv.org/abs/1010.5769}{{\ttfamily arXiv:1010.5769 [math.NT]}}.

\bibitem{cite-key}
S.~C. Rose, ``Quasi-modularity of generalized sum-of-divisors functions,''
  \href{http://dx.doi.org/10.1007/s40993-015-0019-1}{{\em Research in Number
  Theory} {\bfseries 1} no.~1, (2015) 18},
  \href{http://arxiv.org/abs/1506.04963}{{\ttfamily arXiv:1506.04963
  [math.NT]}}.


\bibitem{Hosono:1999qc}
S.~Hosono, M.~H. Saito, and A.~Takahashi, ``{Holomorphic anomaly equation and
  BPS state counting of rational elliptic surface},''
  \href{http://dx.doi.org/10.4310/ATMP.1999.v3.n1.a7}{{\em Adv. Theor. Math.
  Phys.} {\bfseries 3} (1999) 177--208},
  \href{http://arxiv.org/abs/hep-th/9901151}{{\ttfamily arXiv:hep-th/9901151}}.

\bibitem{Huang:2013yta}
M.-x. Huang, A.~Klemm, and M.~Poretschkin, ``{Refined stable pair invariants
  for E-, M- and $[p, q]$-strings},''
  \href{http://dx.doi.org/10.1007/JHEP11(2013)112}{{\em JHEP} {\bfseries 11}
  (2013) 112}, \href{http://arxiv.org/abs/1308.0619}{{\ttfamily arXiv:1308.0619
  [hep-th]}}.


\bibitem{eichler2013theory}
M.~Eichler and D.~Zagier, {\em The Theory of Jacobi Forms}.
\newblock Progress in Mathematics. Birkh{\"a}user Boston, 1985.

\bibitem{Dabholkar:2012nd}
A.~Dabholkar, S.~Murthy, and D.~Zagier, ``{Quantum Black Holes, Wall Crossing,
  and Mock Modular Forms},'' \href{http://arxiv.org/abs/1208.4074}{{\ttfamily
  arXiv:1208.4074 [hep-th]}}.

\end{thebibliography}

\providecommand{\href}[2]{#2}\begingroup\raggedright\endgroup

\end{document}